\documentclass[12pt]{article}
\usepackage{graphicx}
\def\bcc{\begin{center}}
\def\ecc{\end{center}}

\def\dm2{\rm{\Delta m^2}}

\def\s2tw{\rm{ sin ^2 \theta _W }}
\def\am241{\rm{ ^{241} Am }}
\def\u238{\rm{ ^{238} U }}
\def\th232{\rm{ ^{232} Th }}
\def\k40{\rm{ ^{40} K }}
\def\th232{\rm{ ^{232} Th }}
\def\u238{\rm{ ^{238} U }}
\def\cs137{\rm{^{137} Cs }}
\def\ba133{\rm{^{133} Ba }}

\def\s2tw{\rm{ sin ^2 \theta _W }}

\def\ke10{\rm{\kappa_e}}

\begin{document}

\bcc
{\Large Multifragmentation Studies in $^{84}Kr$ Interactions with Nuclear 
Emulsion at around 1 A GeV}
\ecc

\bcc
V. SINGH\footnote{vsingh@phys.sinica.edu.tw}, S.K. TULI\\
Department of Physics, Banaras Hindu University, Varanasi, India\\
B. BHATTACHARJEE, S. SENGUPTA\\
Department of Physics, Gauhati University, Guwahati, India\\
A. MUKHOPADHYAY\\
Department of Physics, North Bengal University, New Jalpaiguri, India\\
\ecc



\bcc
Abstract
\ecc
Projectile fragmentation of $^{84}Kr$ in three 
different energy 
intervals has been studied. Many aspects of multifragmentation
process have been examined in depth. It is observed that 
multifragmentation is a general low energy phenomenon associated
with heavy beam. The number of Intermediate-Mass-Fragments (IMF's)
shows strong projetile mass dependence.

\section{Introduction}

The interactions of relativistic nuclei with target nuclei
have been studied by number of authors using several types
of experimental techniques {\bf[1-2]}. The nuclear emulsion technique
provides a global view of the phenomena a view that can hardly
be matched by other techniques which provide specific data
with higher statistical significance. High energy heavy-ion
collisions offer an unique opportunity to study new phase of
matter. In nuclear collisions at lower beam energies ($<$ 1 A 
GeV), breakup of the interacting nuclei into different fragments
dominates.
A new kind of breakup process has been observed recently in
which many fragments, each more massive than an alpha
particle, are emitted in an interaction and related to the 
production of highly excited nuclear system {\bf [3-6]}. This process is
called multifragmentation because it is characterized by the
production of several medium size, moderatly excited fragments
 {\bf [7]}. Many experimental and theoretical {\bf [8-11]} studies 
have been done,
but the underlying physics is not yet clear due to lack of
experimental data.
In this paper we present results obtained from an experiment
in which nuclear emulsions were exposed to Krypton ($^{84}Kr$)
nuclei with an incident energy of 1 A GeV. We compare our
results with the results of Lanthanum ($^{139}La$), Gold ($^{197}Au$) and 
Uranium ($^{238}U$) at nearly the same energy.

\section{Experimental Details}

In the present experiment, we have employed a stack composed
of {\bf NIKFI BR2} nuclear emulsion pellicles of dimensions
9.8 $\times$ 9.8 $\times$ 0.06 ~$cm^3$, exposed horizontally to $^{84}Kr$
ions at about 1 A GeV at the {\bf SIS} synchrotron at {\bf GSI}, 
Darmstadt (Germany). The events have been examined and analyzed
with the help of a {\bf LEITZ (ERGOLUX)} optical microscope.
In order to obtain an unbaised sample of events, an along-the-track
scanning technique has been employed using an oil immersion objective
of 100X magnification with a digitized microscope readout. The beam 
tracks were picked up at a distance of 4 mm from the edge of 
the plate and carefully followed until they either interacted 
with emulsion nuclei or escaped through any one surface of the 
emulsion or stopped in the plate.

The interaction mean free path of $^{84}Kr$ in nuclear emulsions
has been determined and found to be 6.76$\pm$0.21 (1197 events) cm.
The {\bf DGKLMTV} collaboration {\bf [12]} found a value of mean free
path ($\lambda$) 7.10$\pm$0.14 (877 events) cm, consistent within the
experimental error with our value. The classification of secondary 
charged particles in these events has been reported earlier {\bf [12]}.
As the beam passes through the emulsions, it loses its energy
due to ionization and eventualy stops in the nuclear emulsion.
Since the beam energy decreases as beam go from the entrance 
edge, we have divided each plate in three major regions where
the beam has energy in the range {\bf A} (0.95 - 0.80 A GeV),
{\bf B} (0.80 - 0.50) and {\bf C} below 0.50 A GeV (0.08 -0.50
 A GeV), respectively.

In the following, we will deal only with those interactions in 
which at least one charged particle was produced. Example of
electromagnetic dissociation of target and projectile nuclei
have been rejected. The rest data samples (1100 events) in the 
three energy intervals contain 727, 337 and 36 events, respectively.

The charge associated with each projectile fragments has been
estimated. The charge of light PF's (Z$<$10 charge unit) has 
been measured by blob/gap density and by determining the gap 
length coefficient. For medium PF's (Z up to 19 charge unit)
the charge has been determined by delta-rays ($\delta$) density
measurments {\bf [13]} . For still heavier fragments the assignment of 
charge has been done by estimating the limits on charge from 
the range of the fragments and by comparing the fragment width
relative to the beam track width {\bf [14]}. The charge estimation has 
an accuracy of $\pm$1 charge unit.

\section{Experimental Results and Discussion}

The mean multiplicities of Z$\ge$3 projectile fragment and Z=2 PF
of complete set of 1100 events distributed over the entire
energy range of $^{84}Kr$ beam ($~$950 A MeV - 80 A MeV) are 
1.21$\pm$ 0.04 and 2.03$\pm$ 0.06, respectively. 
\begin{figure}
\center
\includegraphics[height=8cm,angle=0]{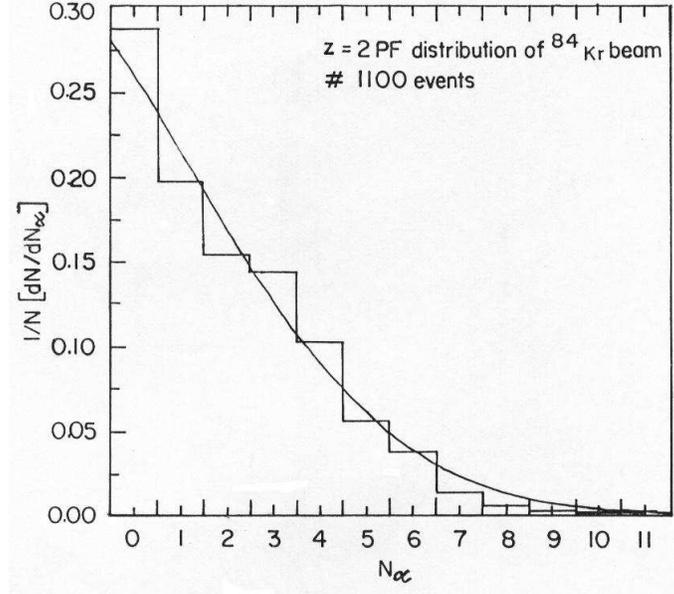}
\caption{
Frequency distribution of Alpha fragments in $^{84}Kr$ interactions
at about 1 A GeV with the Gaussian fit.
}
\end{figure}
The $<N_{f}>$ and  $<N_{\alpha}>$
values in the above mentioned three energy intervals are 1.17 
$\pm$0.05 and 2.02$\pm$ 0.08 (interval {\bf A}), 1.22$\pm$ 0.06 
and 2.04$\pm$ 0.10
(interval {\bf B}) and 2.06$\pm$ 0.35 and 2.37$\pm$ 0.40 (interval {\bf C}), 
respectively. The multiplicity distribution of Z=2 PF of $^{84}Kr$
for complete set of minimum bias events is shown in {\bf Figure 1}.
The distribution admits a Gaussian fit with a width of 8.76 and 
a tail extending upto 11. Events having no Z=2 PF are more
probable than other type of events.

We have calculated for the alpha fragment multiplicity distribution
the $C_{q}$ moments ($C_{q}$ = $<n_{\alpha} ^{q}>$ / $<n_{\alpha}>^{q}$
in the minimum bias sample of interactions of $^{84}Kr$ beam in
the three energy intervals. The values are tabulated in {\bf Table 1}.
\begin{table}
\caption{$C_{q}$ moments of Z=2 PF's from $^{84}Kr$.}
\begin{tabular}{cccc}
\hline
&&Energy intervals&\\
$C_{q}$&{\bf A}&{\bf B}&{\bf C}\\
\hline
$C_{2}$&1.86$\pm$0.07&2.08$\pm$0.10&1.72$\pm$0.26\\
$C_{3}$&4.26$\pm$0.17&5.59$\pm$0.28 &3.42$\pm$0.51\\
$C_{4}$&11.25$\pm$0.45&18.07$\pm$0.90&7.48$\pm$1.12\\
\hline
\end{tabular}
\end{table}
For the same order q these moments show only a week variation with energy
and no definite conclusions are possible in view of limited statistics.
However there is clear variation (increase) in $C_{q}$ with increasing
order q of the moment.

We have plotted $N_{\alpha}$ with respect to the mass number
($A_{p}$) of the source projectile nucleus for the different
minimum biased reactions at nearly similar lab energy in {\bf Figure 2}.
\begin{figure}
\center
\includegraphics[height=8cm,angle=0]{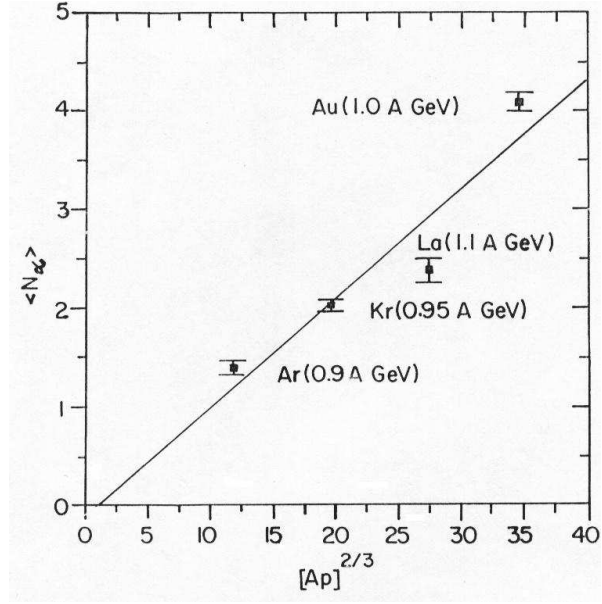}
\caption{
Plot of $<N_{\alpha}>$ Vs $A^{2/3}$ for different projectiles at nearly
similar lab energy. Solid line is the best-fit.
}
\end{figure}
\begin{figure}
\center
{\bf a)}
\includegraphics[height=7cm,angle=0]{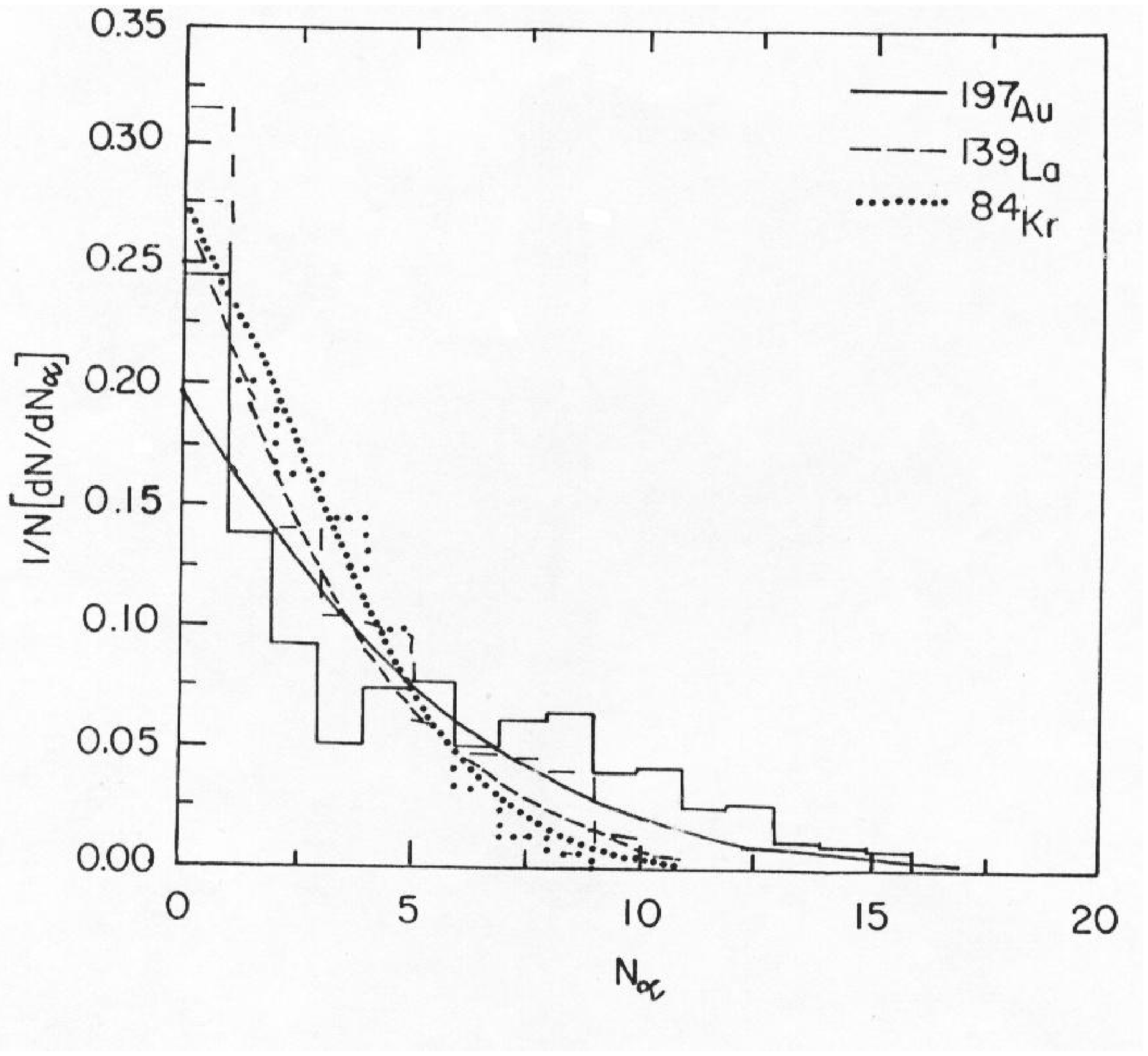}
{\bf b)}
\includegraphics[height=7cm,angle=0]{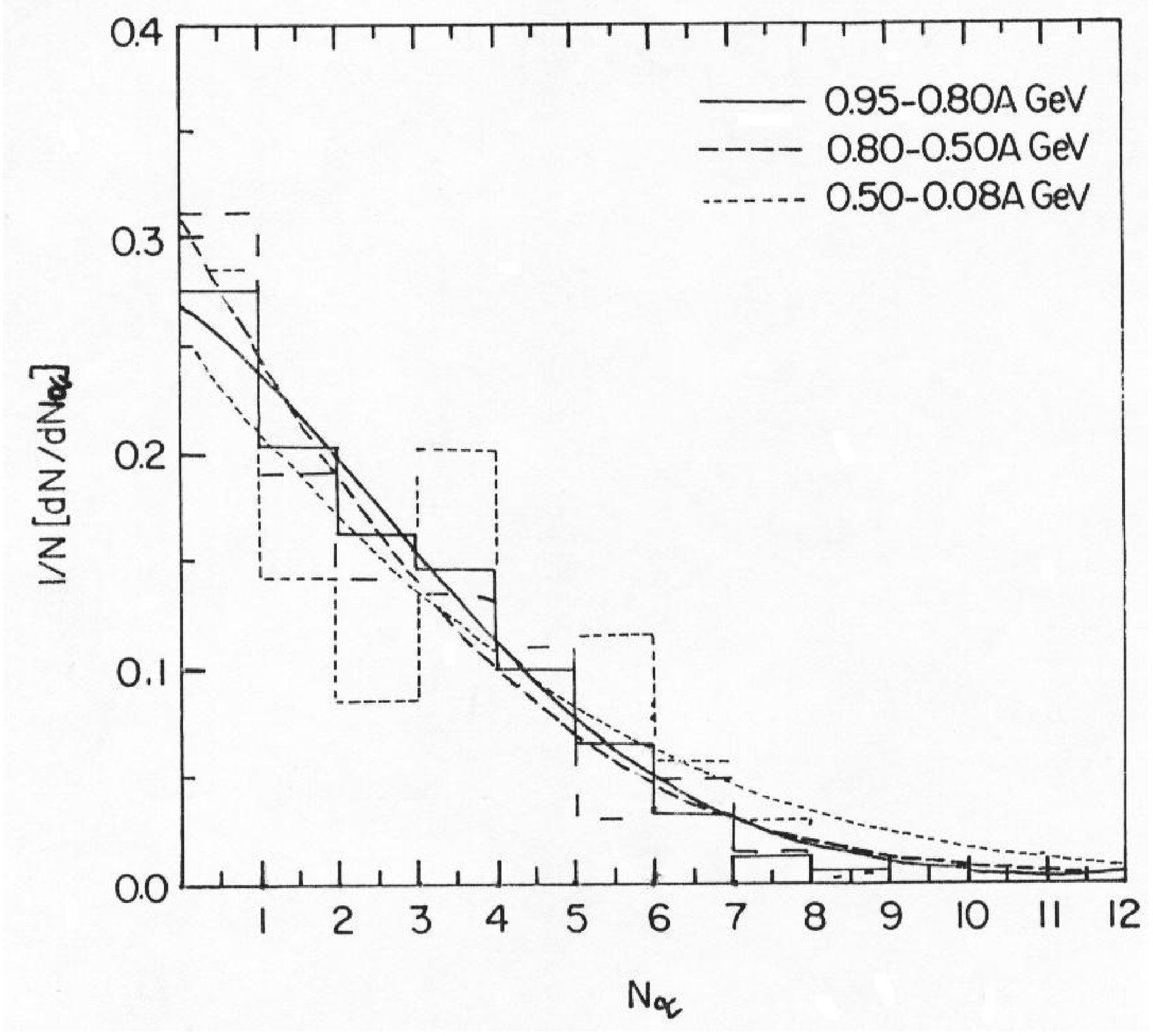}
\caption{
{\bf (a)} Frequency distribution of alpha fragments in $^{84}Kr$, $^{139}La$
and $^{197}Au$ interactions at 0.60 - 0.99 A GeV. Solid, dashed and
dotted lines are Gaussian fits for Au, La and Kr, respectively.
{\bf (b)} Frequency distribution of alpha fragments in $^{84}Kr$ interactions
in different energy intervals. Solid, dashed and dotted lines are
Gaussian fits for Kr in the mentioned energy intervals ({\bf A},
{\bf B} and {\bf C}, respectively).
}
\end{figure}
The data are examined to check whether the emission of alpha
particles is a surface phenomenon. It is observed that alpha-
multiplicity ($N_{\alpha}$) grows slowly as $(A_{p})^{2/3}$ and
the empirical relation $<N_{\alpha}> = (-0.08\pm 0.11) \times (A_{p})^{2/3}$
represents the experimental data points satisfactorily.

The maximum number and average values of the various type of 
projectile fragments are listed in the {\bf Table 2}. Also
included are the values for the case of gold nuclei at nearly 
the same energy {\bf [15]}.
\begin{table}
\caption{Characteristics of $^{84}Kr$ fragmentation at (0.80 - 0.95
A GeV) and (0.50 - 0.70 A GeV) and comparison with $^{197}Au$ at
(0.10 - 1.00 A GeV).}
\begin{tabular}{cccc}
\hline
Sample&$^{84}Kr$(0.80 - 0.95)A GeV&$^{84}Kr$(0.50-0.70)A GeV&$^{197}Au$(0.01-1.08)A GeV {\bf [15]}\\
\hline
$N_{events}$&234&081&360\\
$<N_{p}>$&15.57$\pm$1.09&10.74$\pm$1.18&16.01$\pm$0.89\\
$<N_{\alpha}>$&01.99$\pm$0.14&01.96$\pm$0.22&05.22$\pm$0.20\\
$<N_{f}>$&01.14$\pm$0.08&01.22$\pm$0.13&02.30$\pm$0.08\\
$<N_{p}^{max}>$&36&36&69\\
$<N_{\alpha}^{max}>$&08&09&15\\
$<N_{f}^{max}>$&04&04&07\\
$<Z^{max}>$&15.43$\pm$1.08&20.12$\pm$2.21&44.47$\pm$1.36\\
\hline
\end{tabular}
\end{table}

The number of released projectile protons $(N_{p})$ is defined
from charge conservation by $N_{p} = Z - (2 \times N_{\alpha} + {\sum}Z_{f})$
, where ${\sum}Z_{f}$ is the sum of charges of all $N_{f}$
projectile fragments with charge $Z_{f}\ge3$ and $N_{\alpha}$ 
is the number of alpha particles. The values listed in this table
show that the average number of released projectile protons $(N_{p})$
depends upon the energy of the projectile, and exhibits no 
drastic variation with the mass number of the beam. At low beam
energy the projectile fragmenation does not involve a large number
of singly charged products and as a result the mean value of the
charge of the heaviest fragment increases. The emission of
Intermediate Mass Fragments ({\bf IMF's}) in case of heavy projectile
is more frequent in comparison with that in case of light projectiles.

The maximum number of Z$\ge$3 fragments and Z=2 alphas emitted is
comparable at the two energies but changes when changing the projectile
mass number from $^{84}Kr$ to $^{197}Au$. 

The fractional yield of alphas (Z=2) from $^{84}Kr$ at $\sim$ 1 A GeV is 
shown in {\bf Figure 3(a)}, and has been fitted by a Gaussian function.
For 1100 minimum biased events the maximum number of alphas in an
event ($N_{\alpha}^{max.}$) and the width of the distribution are 
found to be 11 and 7.88, respectively. The average number of alphas
$<N_{\alpha}>$ is (2.0.$\pm$0.06). For comparison, the data from $^{139}La$
+Em. and $^{197}Au$+Em. at nearly the same energy (0.60-0.99 A GeV) are also
included in the figure. There is significant growth in the number of 
alphas with increasing projectile mass number. The width of the alpha
yields from La and Au are 17.88 and 16.18, respectively.

\begin{figure}
\center
{\bf a)}
\includegraphics[height=7cm,angle=0]{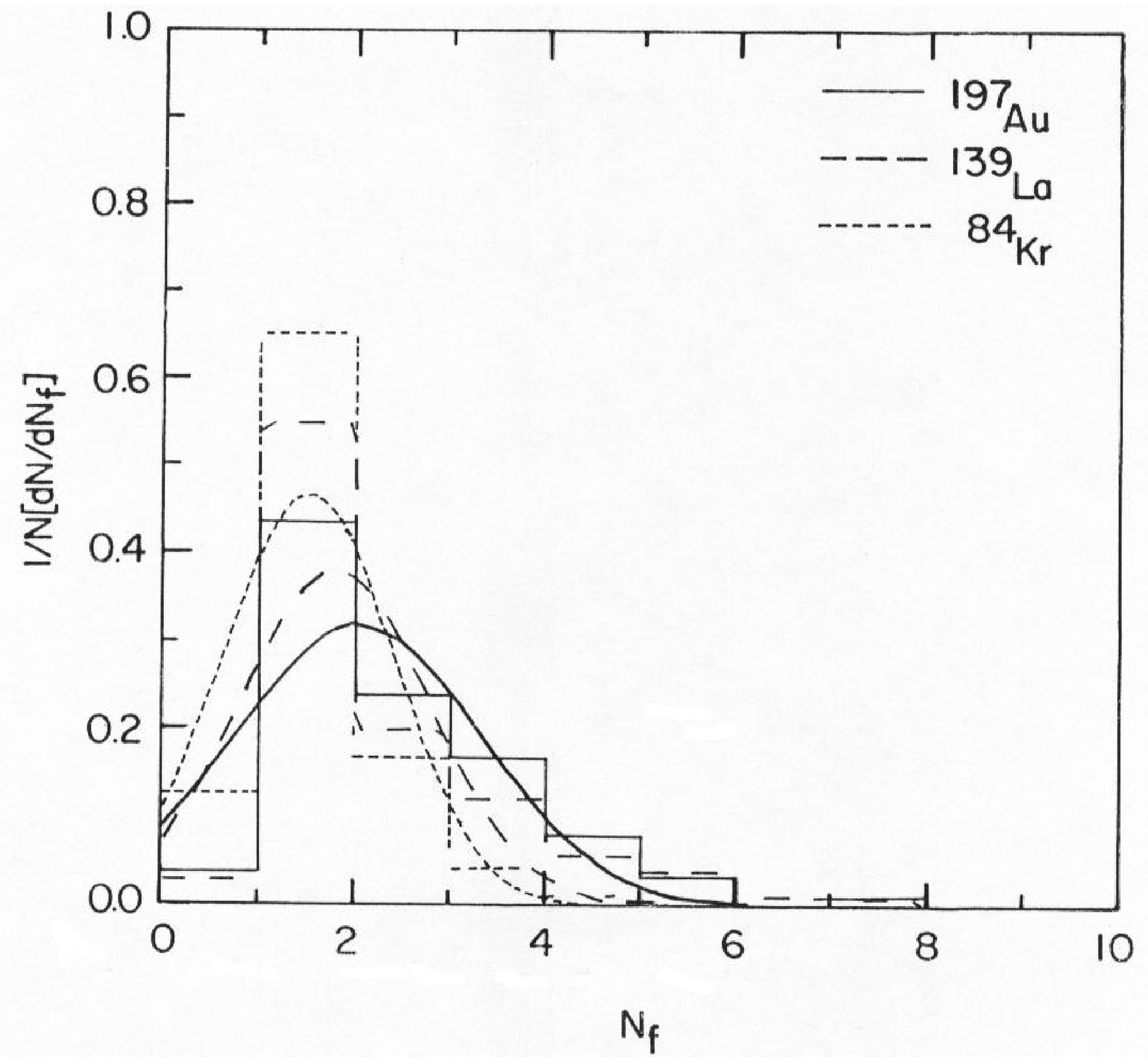}
{\bf b)}
\includegraphics[height=7cm,angle=0]{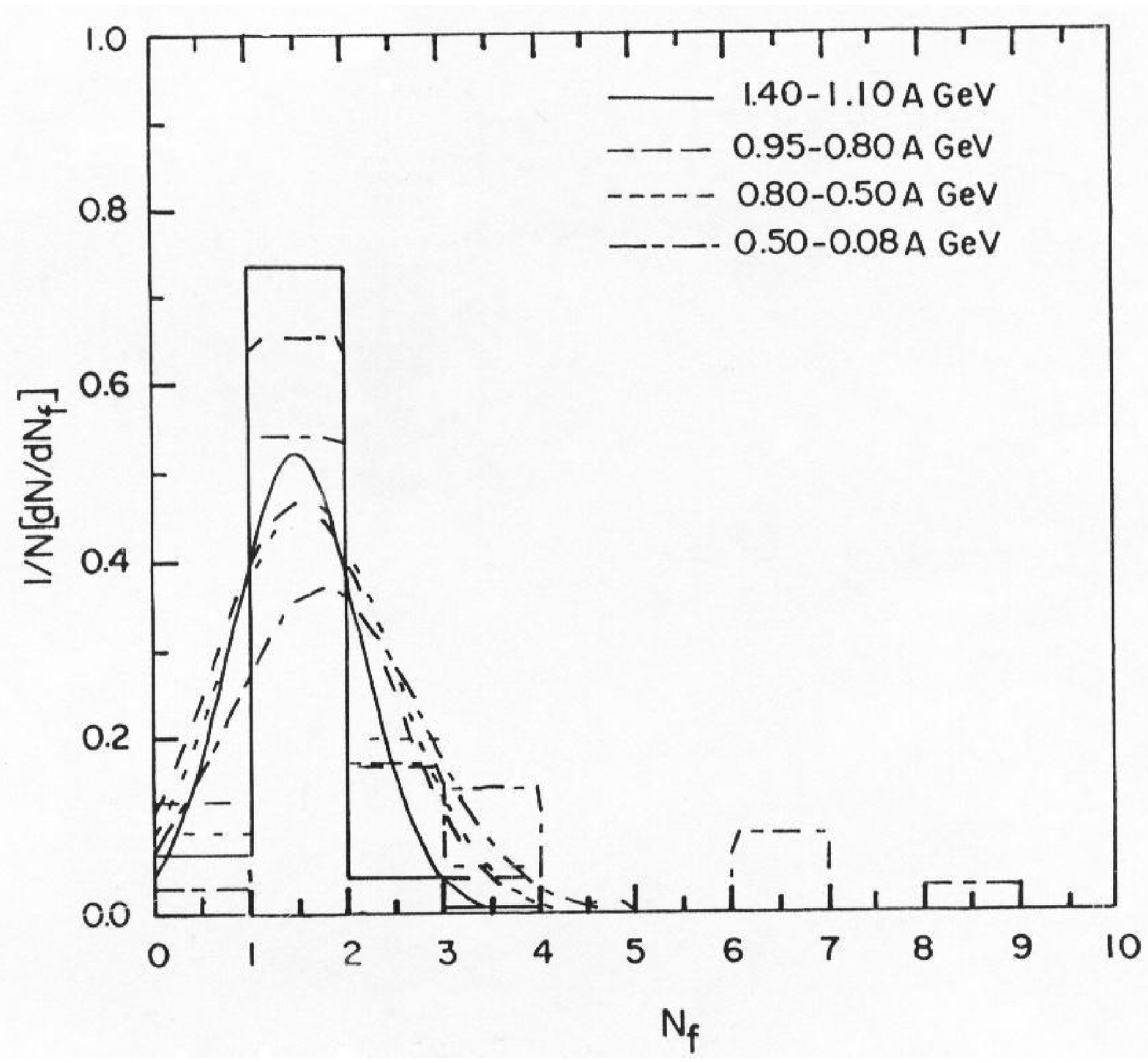}
\caption{
{\bf (a)} Frequency distribution of heavier fragments (Z$\ge$3) from different
beams at 0.60 - 0.99 A GeV together with the Gaussain fit.
{\bf (b)} Frequency distribution of heavier fragments (Z$\ge$3) from $^{84}Kr$ 
interactions in different energy intervals together with the Gaussian fit.
}
\end{figure}
\begin{figure}
\center
{\bf a)}
\includegraphics[height=7cm,angle=0]{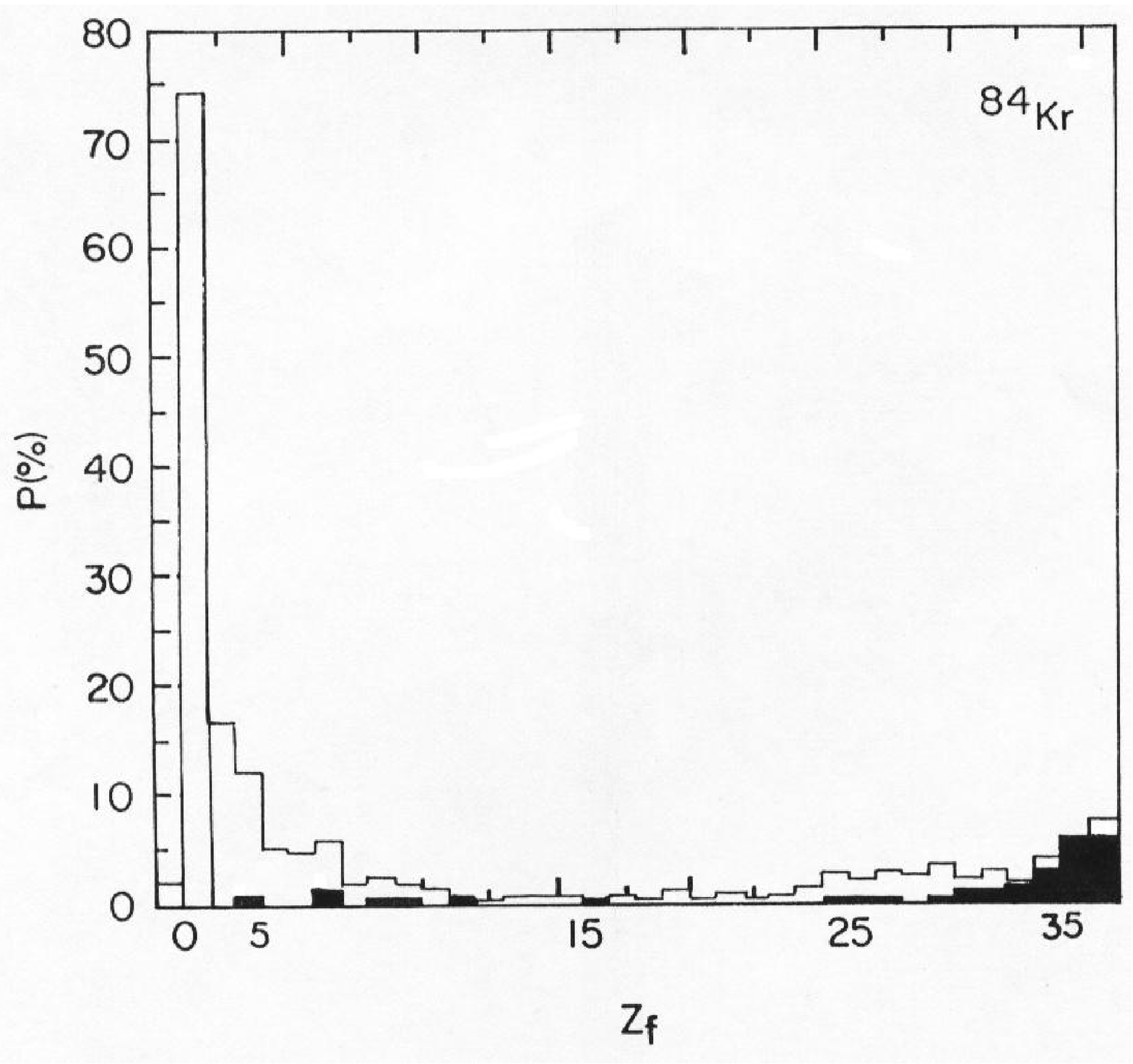}
{\bf b)}
\includegraphics[height=7cm,angle=0]{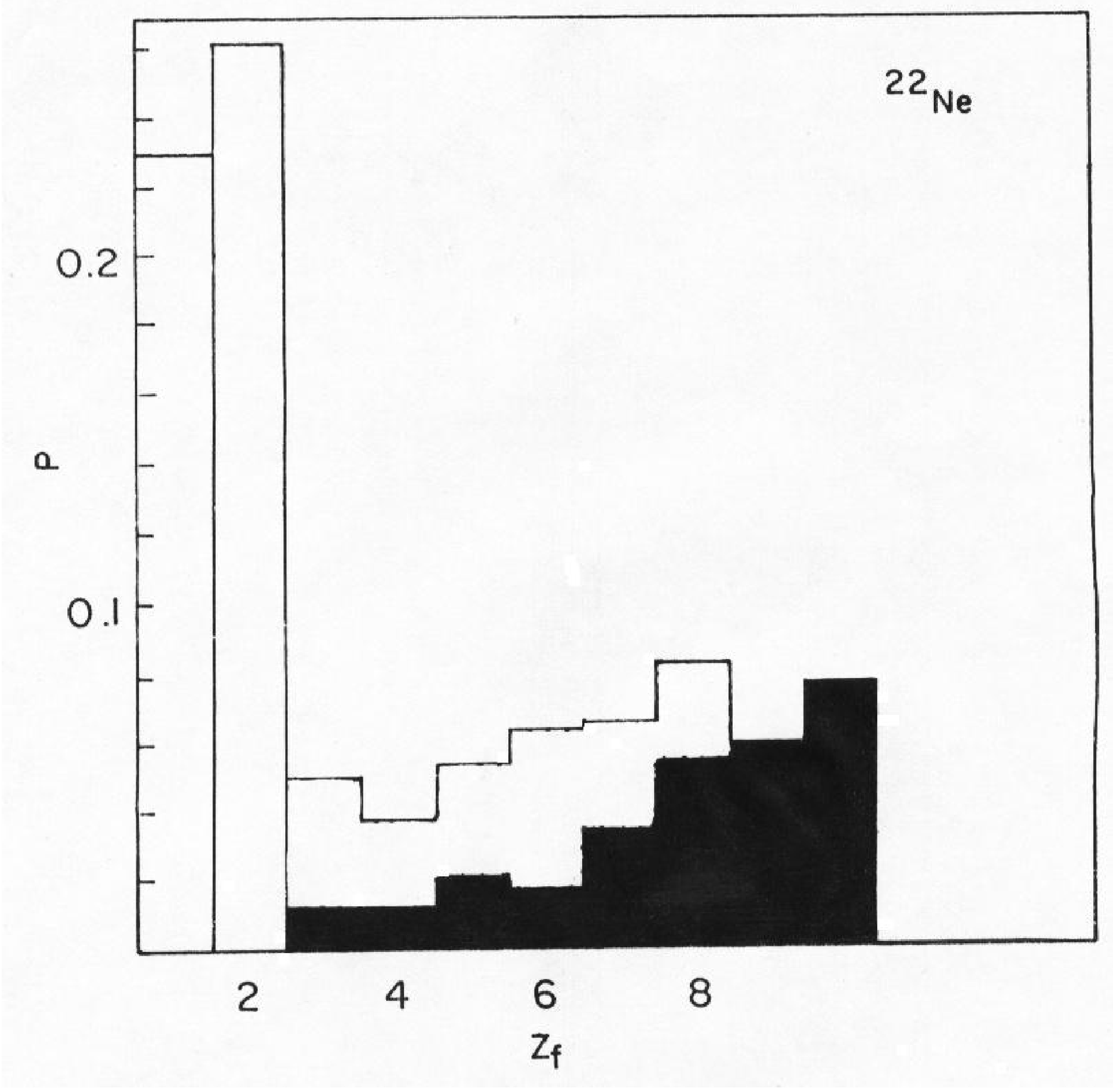}
\caption{
The probability of fragments of {\bf (a)} $^{84}Kr$ and {\bf (b)}
 $^{22}Ne$ projectiles into fragments with charge $Z_{f}$. Solid area
represents the probability for a given fragment $Z_{f}$ not to be 
accompained by alpha particle(s). $Z_{f}$=1 denotes events without alpha
particle(s) or heavier fragment(s).
}
\end{figure}
The fractional yield of alpha particles of $^{84}Kr$ in the three 
energy intervals is shown in {\bf Figure 3(b)}. The three data sets
have been fitted with the Gaussian distribtuion. The widths of the
distributions increase with decreasing beam energy and have a magnitude
of 7.88, 11.04 and 18.12, respectively. The data presented in 
{\bf Figure 3(b)} show that multi-alpha emission is generaly more 
probable at low energy {\bf [16]}.

The yield of Z$\ge$3 projectile fragments from $^{84}Kr$ is depicted
in {\bf Figure 4(a)}, which also includes data from $^{139}La$+Em. and 
$^{197}Au$+Em. for comparison. These distributions have been fitted with 
the Gaussian function. From the $^{84}Kr$ data it is observed that in $~2\%$ 
events $N_{f}$=0 and at same time $N_{\alpha}$=0 while in $~12\%$ events
having $N_{f}$=0 but $N_{\alpha}\ne0$. Heavy beams rarely yield events having
no heavier fragment. The number of events having a single (Z$\ge$3)
fragment increases with decreasing beam mass number.

The frequency spectra of Z$\ge$3 fragments from $^{84}Kr$ beams in the three 
different energy intervals together with the higher (1.4 to 1.1 
A GeV) energy published data {\bf [23]} are shown in {\bf Figure 4(b)} and 
have been 
fitted by Gaussian distributions. In the total sample of 1100 events in 
present experiment $<N_{f}>$=1.21$\pm$0.04. The fitted distribution peaks 
at 1.6 with a width of 1.8. Notice that the maximum number of fragments
in an event is four at high energy (1.4 - 1.1 A GeV) which increases to 
eight at low energy interval({\bf C}) marking a change in the fragmentation
mode.

\begin{figure}
\center
{\bf a)}
\includegraphics[height=6.7cm,angle=0]{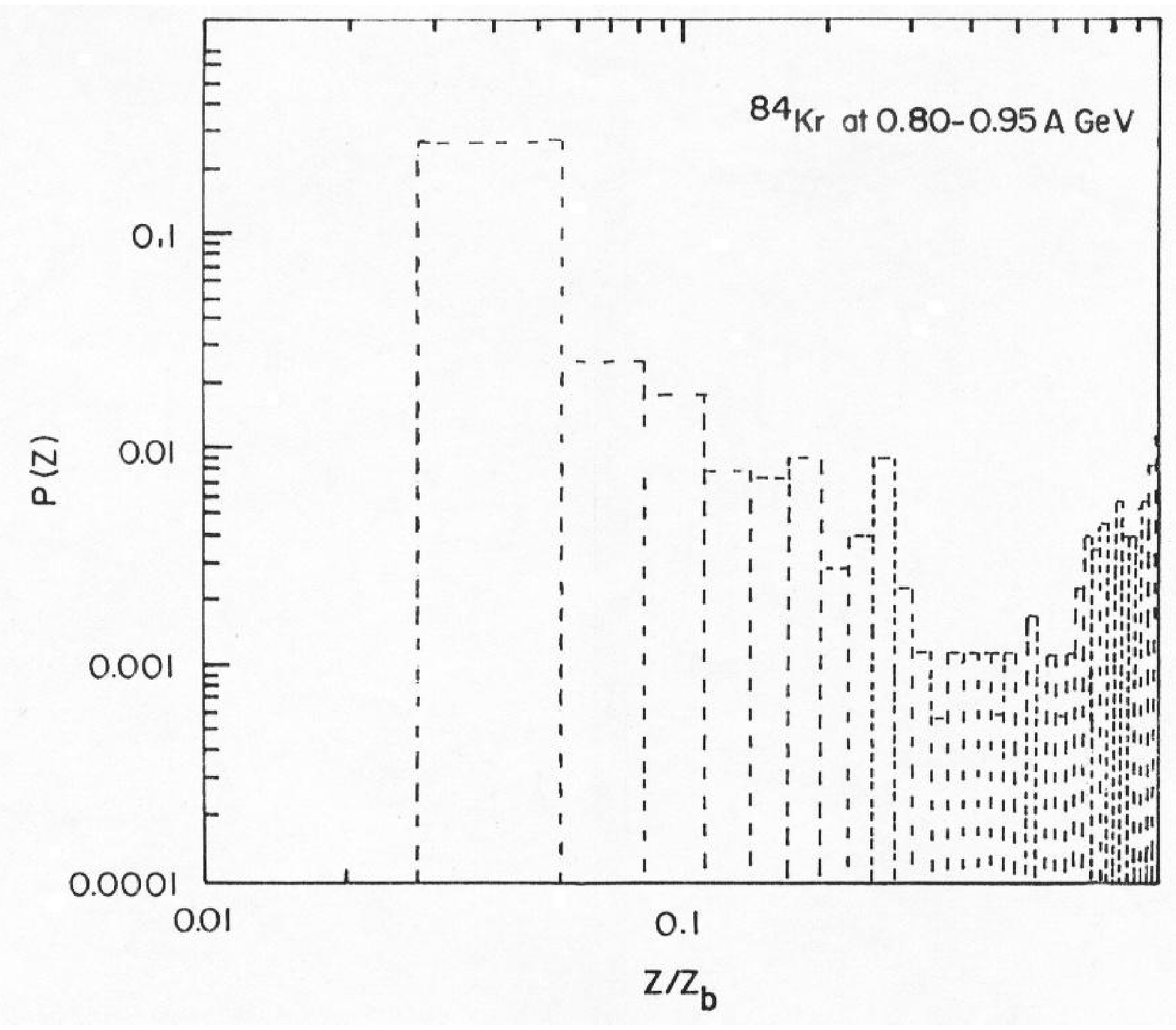}
{\bf b)}
\includegraphics[height=6.7cm,angle=0]{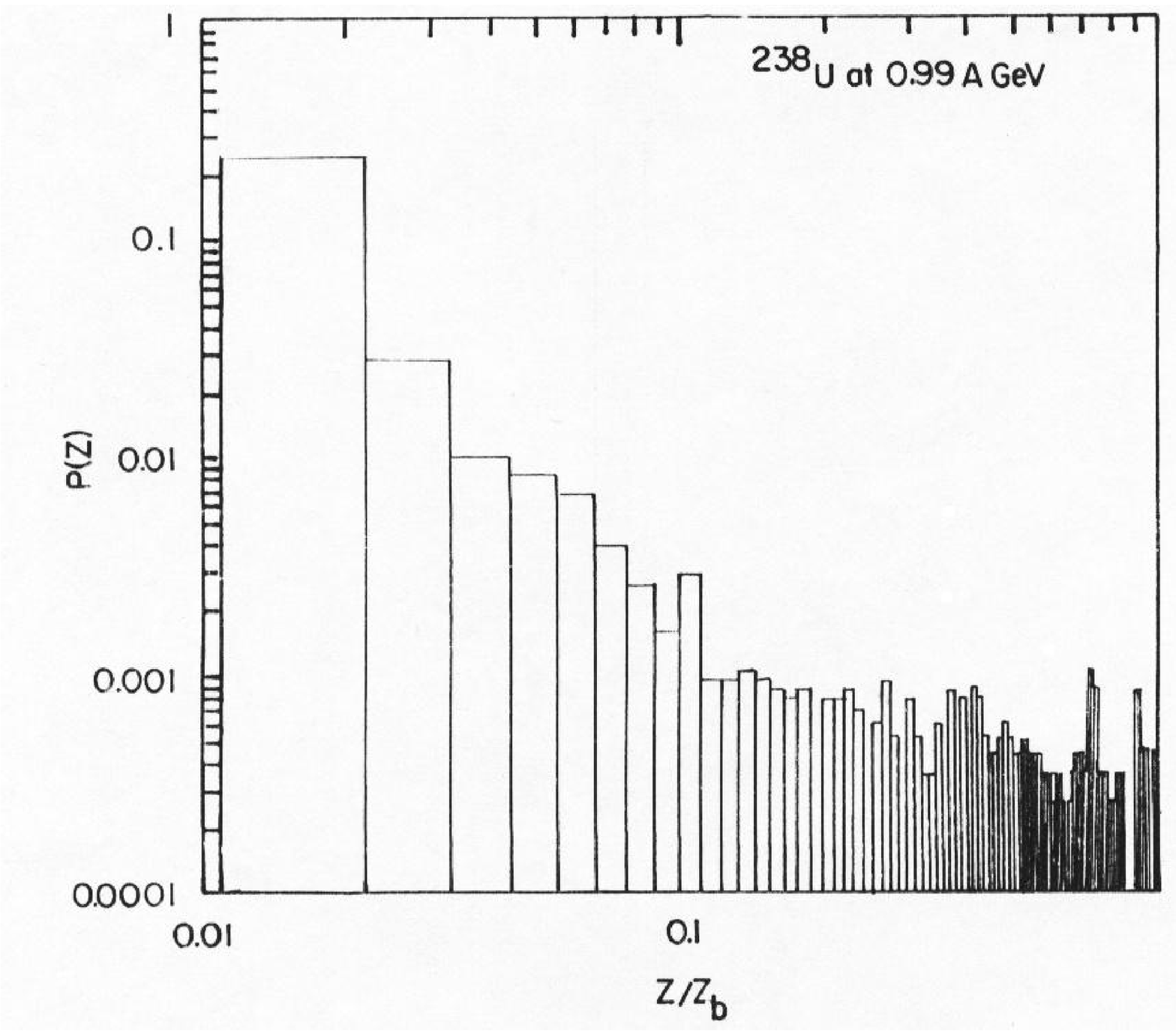}
\caption{
The probability of fragment charge normalised to the beam charge in
{\bf (a)} $^{84}Kr$+Em. and {\bf (b)} $^{238}U$+Em. events at $\sim$
1 A GeV.
}
\end{figure}
\begin{figure}
\center
{\bf a)}
\includegraphics[height=8cm,angle=0]{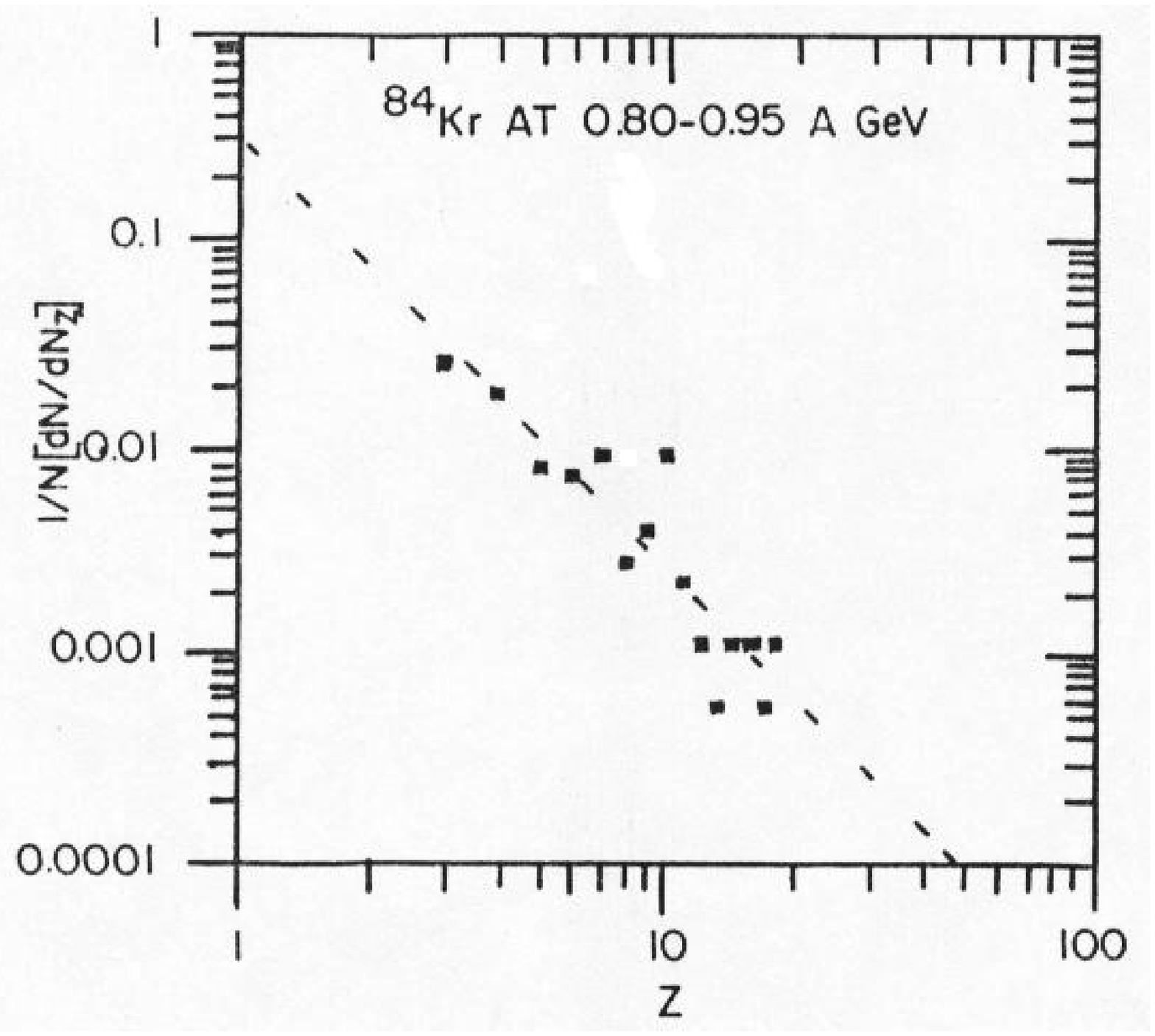}
{\bf b)}
\includegraphics[height=8cm,angle=0]{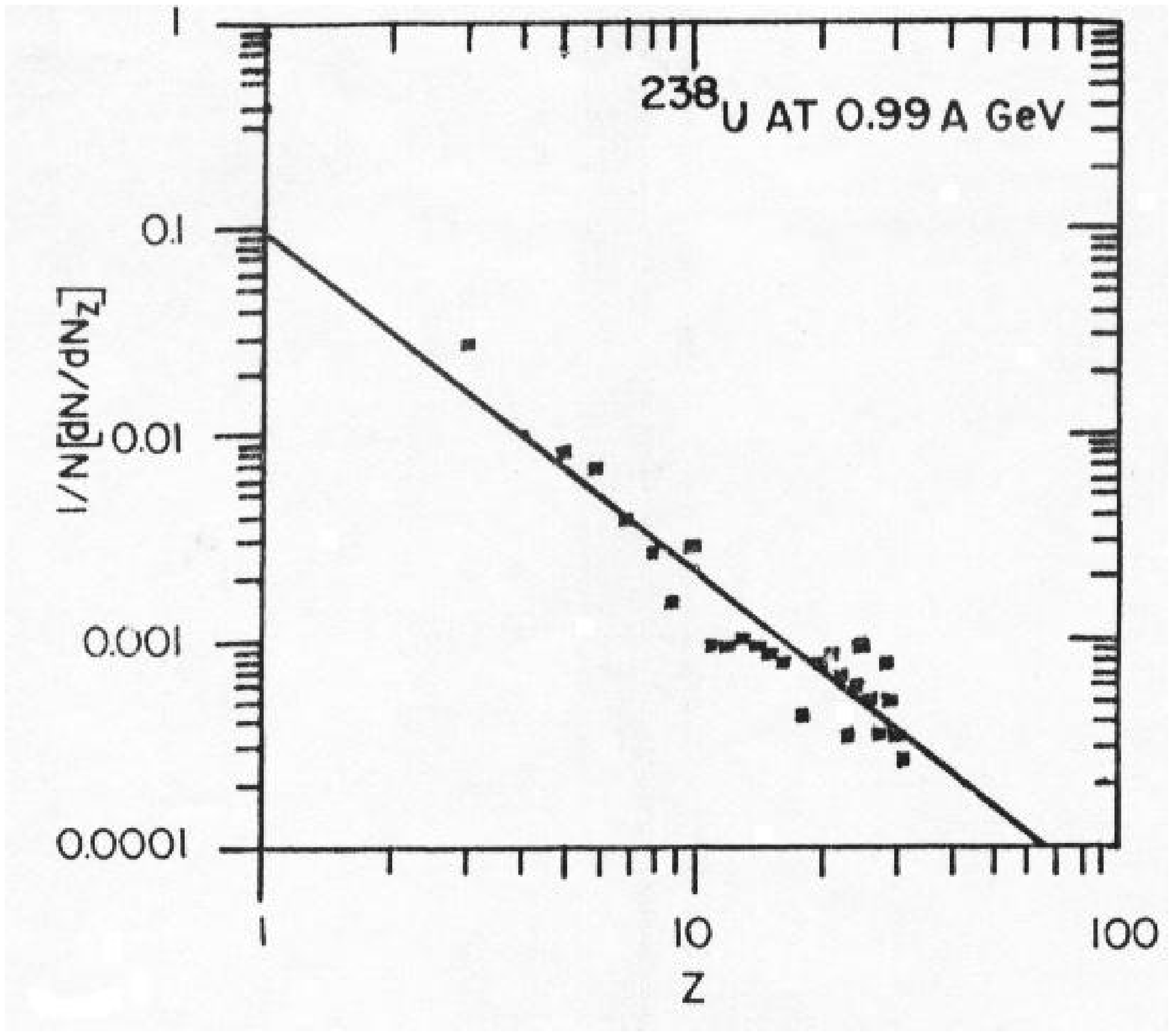}
{\bf c)}
\includegraphics[height=8cm,angle=0]{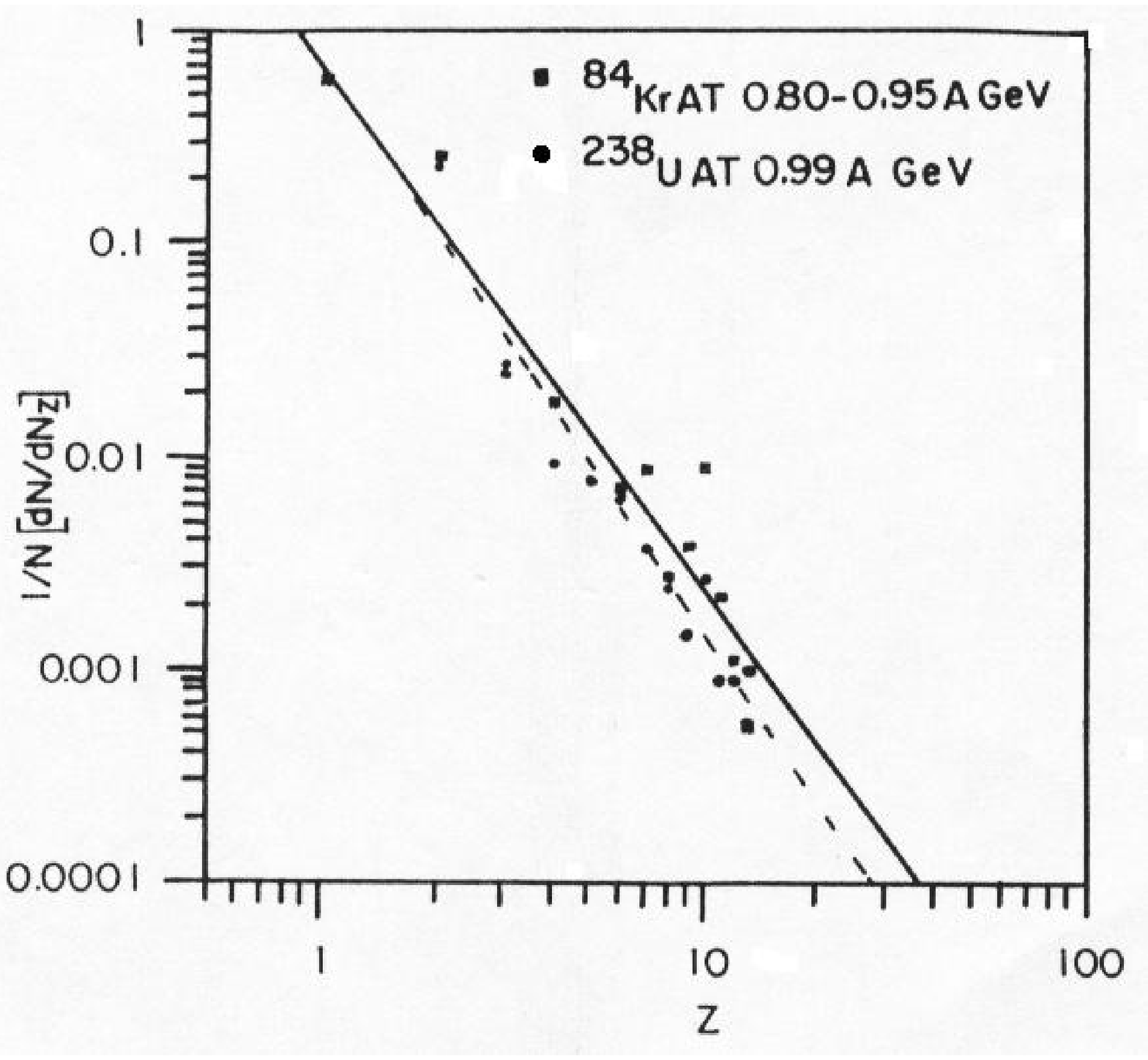}
\caption{
The charge of the projectile fragment fitted with the linear function
(M $\times$ Z + c) for {\bf (a)} $^{84}Kr$ at 0.80 - 0.85 A GeV  and 
{\bf (b)} $^{238}U$ at 0.99 A GeV from Z=3 upto $Z_{b/2}$ and {\bf (c)} 
$^{84}Kr$ at 0.80 - 0.95 A GeV and $^{238}U$ at 0.99 A GeV from Z=1 
upto $Z_{b/2}$.
}
\end{figure}
The probability of multifragmentation (events having two or more fragments 
with Z$\ge$3) in case of $^{84}Kr$ is about $21\%$. It is clear from the    
{\bf Figure 4(a)} that the multifragmentation increases as we go from
$^{84}Kr$ to $^{197}Au$. In other words the multifragmentation is a more
common phenomenon in heavier mass beams. A Gaussian shape fit has been made
to the multiplicity distributions of projectile fragments. The width in 
these fits are broader for heavier mass beam and center of these distributions
are shifted towards higher $N_{f}$. These observations are consistent with
the features expected in multifragmentation process.

\begin{table}
\caption{Gaussian distribution parameters for $^{84}Kr$+Em. in different beam
energy intervals.}
\begin{tabular}{ccc}
\hline
Energy (A GeV)&Center&Width\\
\hline
1.46 - 1.10&1.48&1.33\\
0.95 - 0.80&1.52&1.78\\
0.80 - 0.50&1.60&1.76\\
0.50 - 0.08&1.77&1.96\\
\hline
\end{tabular}
\end{table}
The Gaussian distribution parameters obtained in different beam energy
intervals are summarised in {\bf Table 3}. One can conclude from this table
that with increasing beam energy the width of the distribution decreases
and the location of the peak shifts to lower $N_{f}$ values. The tail of the
distribution extends further for low beam energy. This indicates that the 
energy is more uniformly distributed over a large cross-sectional area in 
low beam energy showing that the multifragmentation is primarily a low 
beam energy phenomenon.

We have already reported the data on projectile fragmenattion for $^{84}Kr$
as a function of beam energy {\bf [17-19]} and have observed that the number of PF's 
with charge one $\&$ two increases with increasing beam energy. But at CERN 
energy, data on PF's from $^{16}O$ show that the yield of fragments 
(Z$\ge$3) increases significantly as the beam energy increases {\bf [20]}.

\begin{table}
\caption{Some parameters describing the projectile fragmentation
for inelastic interactions of $^{22}Ne$ {\bf [20]} and $^{84}Kr$ in 
emulsion.}
\begin{tabular}{cccc}
\hline
&$^{22}Ne$(4.2)&$^{84}Kr$(0.95-0.80)&$^{84}Kr$(0.80-0.08)\\
&A GeV&A GeV&A GeV\\
Fragmentation Mode&Probability (in percents)&for different&fragmentation\\
&modes&&\\
\hline
Two fragments with Z$\ge$3&1.0$\pm$0.2&16.78$\pm$0.67&18.01$\pm$0.90\\
One fragments with Z$\ge$3&50$\pm$1&65.06$\pm$2.60&36.02$\pm$1.80\\
Fragment(s) with Z$\ge$3&29$\pm$1&26.13$\pm$1.05&29.84$\pm$1.49\\
and no alpha particle&&&\\
Fragment(s) with Z$\ge$3&21$\pm$1&61.21$\pm$2.45&61.56$\pm$3.08\\
and alpha particle(s)&&&\\
Alpha particles and no&26$\pm$1&11.28$\pm$0.45&5.43$\pm$0.34\\
heavier fragments&&&\\
No multiply charged&23$\pm$1&1.38$\pm$0.06&2.15$\pm$0.11\\
fragment&&&\\
\hline
&Mean number of&multicharge&fragments per\\
&event&&\\
\hline
Fragments with Z$\ge$3&0.50$\pm$0.01&1.05$\pm$0.04&1.02$\pm$0.05\\
Alpha particles&0.72$\pm$0.01&2.98$\pm$0.12&3.35$\pm$0.17\\
\hline
&Mean charge of&the fragment&\\
\hline
Fragments with Z$\ge$3&7.8$\pm$0.1&14.48$\pm$0.87&17.53$\pm$1.75\\
and no alpha particles&&&\\ 
Fragments with Z$\ge$3&5.4$\pm$0.1&06.53$\pm$0.26&07.96$\pm$0.48\\
and alpha particle(s)&&&\\
\hline
\end{tabular}
\end{table}
The probability distribution of heavy projectile fragments and alpha 
particles emitted from $^{84}Kr$ at around 1 A GeV beam energy is 
presented in {\bf Figure 5(a)}. For comparison with this data, we 
have included a large data set of interactions of $^{22}Ne$ nuclei 
with an energy of 4.2 A GeV in {\bf Figure 5(b)} {\bf [21]}.

We have tabulated some important parameters describing the projectile 
fragmentation of $^{84}Kr$ and compared with the data of $^{22}Ne$ at 
4.2 GeV/nucleon in {\bf Table 4}. From this table a number of 
observations are possible.

The probability of producing one and two fragments is strongly dependent 
on beam mass number. An increase in the probability of producing fragment(s) 
is observed as the mass of the projectile increases. The probability of 
producing fragment(s) with no accompanying alpha particles is independent 
of mass and energy of the beam.

The probability of fragmentation of projectile into alpha particle(s) 
shows strong dependence on the mass of the beam and is independent of 
energy. The probability that only alpha particles but no heavier 
fragments are produced decreases with increasing mass of the projectile. 
The low and high energy $^{84}Kr$ data have a strong dependence on the 
engry of the beam i.e., probability of producing only alpha particles 
increases with increasing beam energy. There is a rapid decrease in 
the number of events with no multiple charged fragments, as the mass 
of the projectile increase.

The mean number of fragments and alpha particles as well as the mean 
charge of the fragments all increase with increasing mass of the 
projectile. The probability of producing multiple fragments with 
Z$\ge$3 increases rapidly with increasing mass of the projectile. 
The mean charge of the fragments with no accompanying alpha particles 
shows rapid increase with increasing mass of the beam. A similar 
feature can be observed  in fragments with alpha particles 
fragmentation mode, which is consistent with the interactions of 
low energy $^{197}Au$ nuclei in emulsions {\bf [22]}.

In the strict thermodynamic sense phase transitions are only defined 
for infinite system. In nature there are a few mesoscopic system with 
numbers of constituents on the order of $10^{2}$ to $10^{5}$. 
Fragmentation of atomic nuclei is a system where we expect to 
experience phase transitions under certain conditions and for which 
mesoscopic finite-size effect should play very important roles. For 
now more than two decades, there have been speculations that 
we may be able to see a first-order phase transition between the 
Fermi liquid of graund state nuclei and the hadronic gas phase of 
individual nucleons and / or small clusters {\bf [23]}. One interesting 
thing is that this first-order transition will terminate at a critical 
point, where the transition becomes continuous and critical exponents 
of nuclear matter can be determined experimentally following the 
so-called Fisher droplet model {\bf [24]}. In this context, particular 
attention must be focus on the power-law dependence of the yield of 
nuclear fragments as a function of mass number.

The charge distribution of projectile fragments from $^{84}Kr$ 
interactions is shown in {\bf Figure 6(a)} and the corresponding 
distribution from $^{238}U$ is included as {\bf Figure 6(b)} for 
\begin{figure}
\center
{\bf a)}
\includegraphics[height=7cm,angle=0]{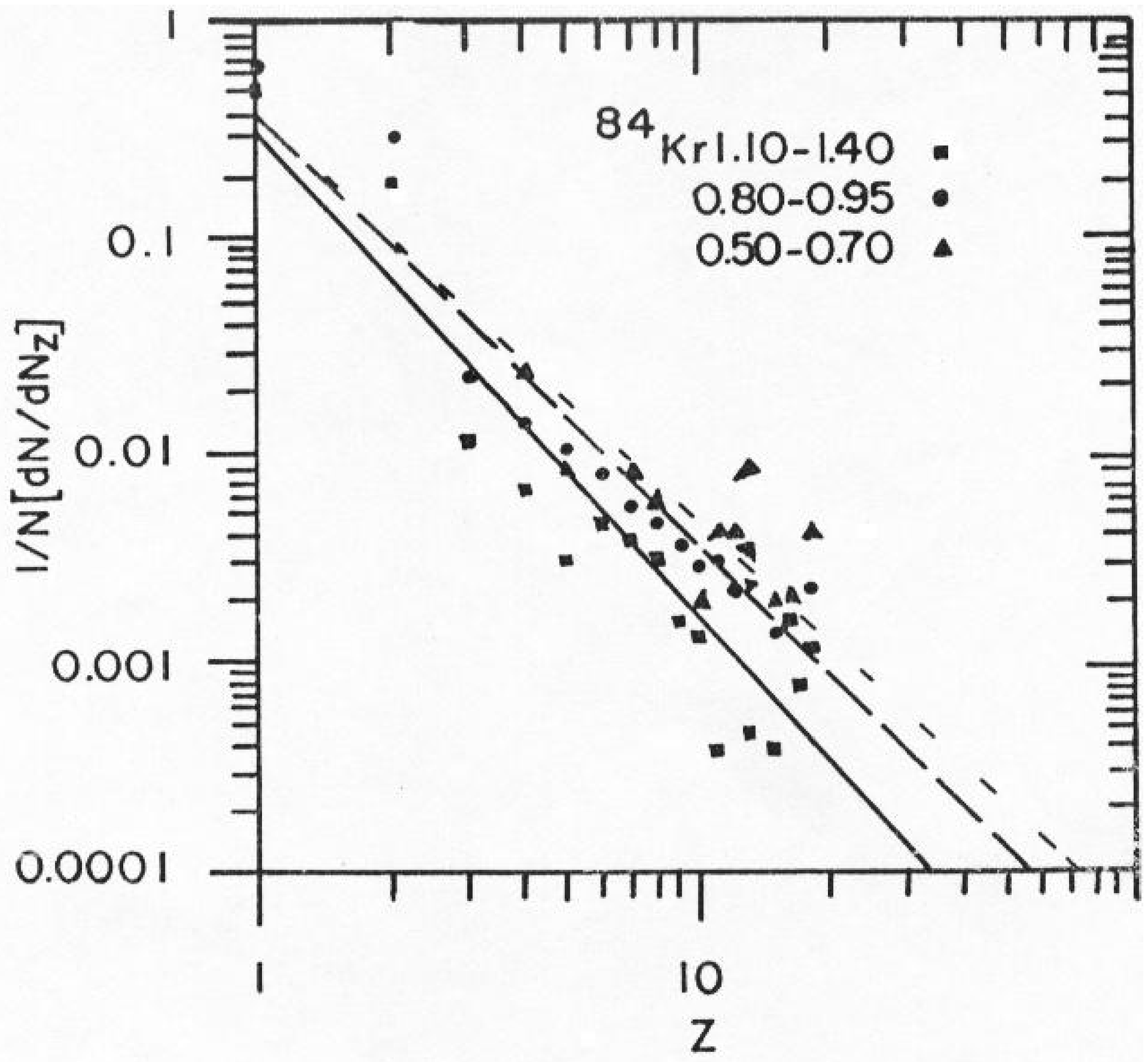}
{\bf b)}
\includegraphics[height=7cm,angle=0]{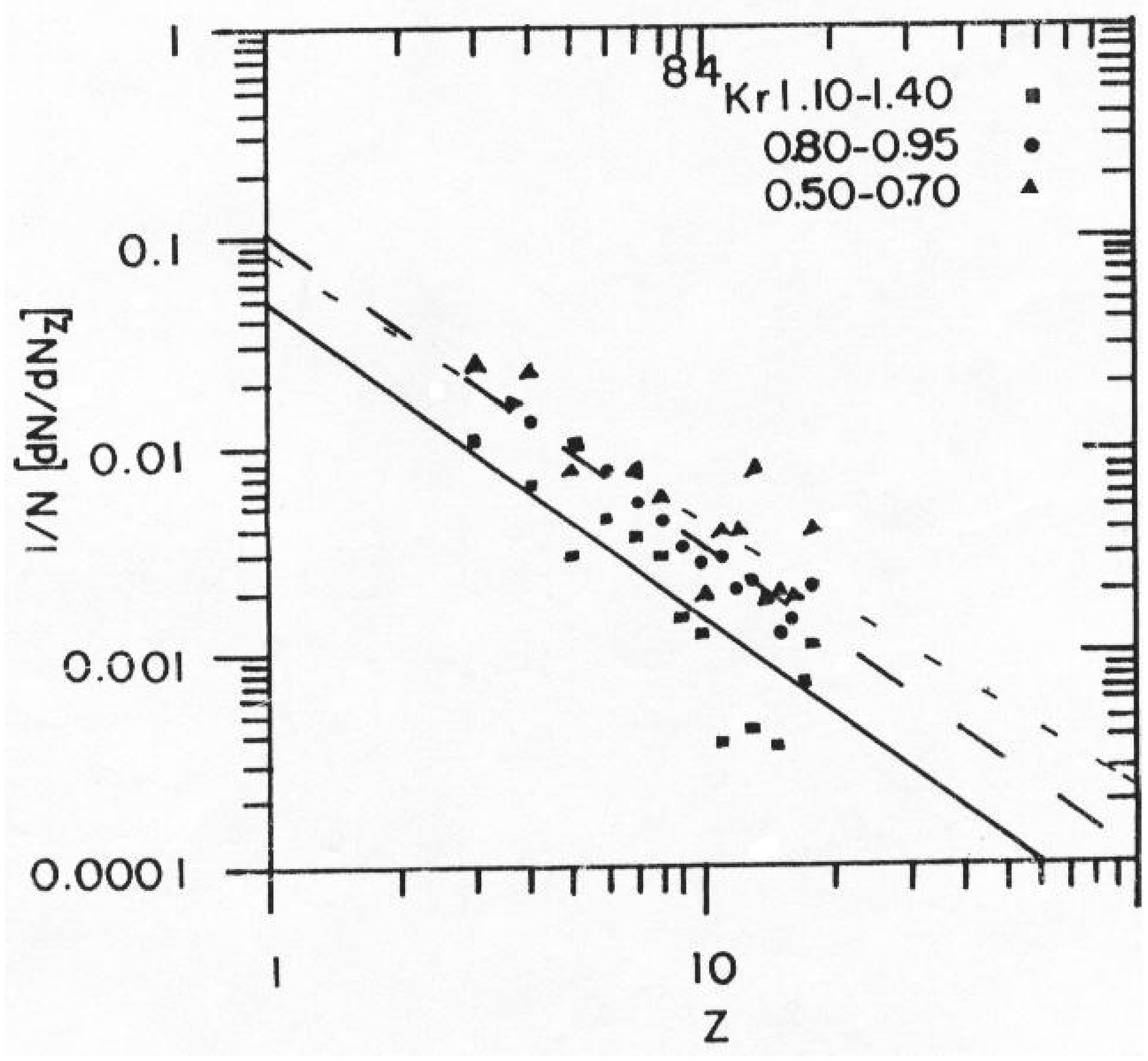}
\caption{
The charge of the projectile fragment fitted with the linear function
{\bf (a)} from 1 to $Z_{b/2}$ and {\bf (b)} from 3 to $Z_{b/2}$ for
$^{84}Kr$ in energy intervals 1.1 - 1.4, {\bf A} and 0.50 - 0.70 A GeV.
}
\end{figure}
comparison. The figure shows as increase in the yield of low 
charge PF's in heavier beams. This indicates a greater degree of 
breakup for heavier beams {\bf [25]}. To facilitate inter comparison of 
fragment spectrum from beams of different charge we choose to 
show the distribution as a function of the charge of the fragment 
divided by the beam charge. Clearly, in case of $^{84}Kr$ we can see 
the U-shaped mass yield spectrua and the power-law dependence of the 
yield of the low-to heavier mass fragments upto the mass.

The charge distribution of all the projectile fragments produced 
in our interactions are shown in the {\bf Figure 7(a)}. An inverse 
power law given by f(Z) ${\alpha}$ Z$^{\tau}$ reproduces the major 
features of this distribution remarkably well. The charge spectrum
 from Z=3 up to the $Z_{b}$/2 has been fitted with the above 
function and compared with $^{238}U$ in {\bf Figure 7(b)}. 
{\bf Figure 7(c)} represents the data for $^{84}Kr$ and $^{238}U$ 
from Z=1 to $Z_{b}$/2. The values of the exponents for $^{84}Kr$ 
are -2.45$\pm$0.16 and -1.49$\pm$0.09 for charges 1 to $Z_{b}$/2 
and 3 to $Z_{b}$/2, respcetively. Notice, this simple trend is 
not exhibited by PF of excessively large charge {\bf [26]}. The rise in 
the yield of heavier fragments is due to the events in which only 
a single heavy fragment Z$\ge$3 is emitted.

\begin{table}
\caption{The value of the exponent for different beams at nearly same
($\sim$1.0 A GeV) energy.}
\begin{tabular}{ccc}
\hline
Beam&Z from 1 to $Z_{b/2}$&Z from 3 to $Z_{b/2}$\\
&Exponent&Exponent\\
\hline
$^{238}U$&-(2.67$\pm$0.12$)^{*}$&-(1.66$\pm$0.07)\\
$^{197}Au$&-(2.55$\pm$0.08)&-(1.56$\pm$0.05)\\
$^{84}Kr$&-(2.45$\pm$0.16)&-(1.49$\pm$0.09)\\
\hline
* The exponent ($\tau$) is -(2.01$\pm$0.09) up to Z=25 {\bf [27]} 
\end{tabular}
\end{table}
\begin{table}
\caption{The value of the exponent of $^{84}Kr$ beam in different 
energy intervals.}
\begin{tabular}{ccc}
\hline
Energy (A GeV)&Z from 1 to $Z_{b/2}$&Z from 3 to $Z_{b/2}$\\
&Exponent&Exponent\\
\hline
0.84 - 1.52&-(2.29$\pm$0.09$)^{*}$&-(1.50$\pm$0.06)\\
0.80 - 0.95&-(2.05$\pm$0.13)&-(1.49$\pm$0.09)\\
0.50 - 0.70&-(1.90$\pm$0.21)&-(1.27$\pm$0.14)\\
\hline
* The exponent ($\tau$) is -(2.91$\pm$0.11) up to Z=11 {\bf [28]} 
\end{tabular}
\end{table}
To compare the results from different beams at the same energy and 
those from $^{84}Kr$ beam in different energy intervals we summarise 
the fit parameters in {\bf Tables 5 and 6}. On examining the data from 
different beams at nearly the same beam energy, it is found that the 
values of the exponent increase weakly with increasing mass of the 
projectile. This holds also for the fit from 3 to $Z_{b}$/2. It shows 
that the emission of intermediate mass fragment {\bf (IMF)} increases 
with increasing the mass of the projectile.

The values of the exponent at different beam energies of ~$^{84}Kr$, 
as plotted in {\bf Figure 8 (a and b)}, tabulated in {\bf Table 6}, show an 
increasing trend with beam energy. An increase in the slope of the 
fitted straight lines with increasing beam energy has earlier 
been reported {\bf [15]} .

An inverse power law dependence of mass yield has been interpreted as 
a signature for the occurence of statistical clustering expected in 
the liquid - gas phase near its critical point {\bf [29,24]} and this power - 
law is similar to the droplet size distribution close to the critical 
temperature in the Fisher's theory of condensation {\bf [30]}. Power - law 
behaviors have also been seen in cluster - size distributions at 
the percolation threshold {\bf [31,32]}. Such a power - law shows the validity 
of a liquid - gas phase transition at a critical temperature.

\section{Conclusions}

The following conclusions have been drawn from the results of the 
present work and the subsequent analysis.

The $<N_{p}>$ is strongly dependent on the energy of the projectile. 
It is a good measure of the total excitation energy introduced in 
the system. The average value of the charge carried by the heaviest 
projectile fragment in an event is more at low projectile energy. 
The number of IMF's shows strong projectile mass dependence.

There is a systematic increase in  $<N_{\alpha}>$ and  $<N_{f}>$ 
with the increase in projectile mass.

Observations indicate that the multifragmentation is a more common 
phenomenon in heavier mass beams and at lower beam energy. The
 energy distribution over large volume is more uniform at low 
beam energy.

The charge distribution of projectile fragments shows an inverse 
power law [ f(Z) ${\alpha} ~ Z^{\tau}$] behaviour. At the 
same energy, the heavier beams undergo greater degree of breakup. 
Power - law dependence is consistent with the validity of liquid 
- gas phase transition at this energy.

\section{Acknowledgments}

Partial support from DAE (Department of Atomic Energy) and UGC 
(University Grant Commission), Govt. of India for carrying out 
this work is gratefully acknowledge.


\end{document}